\documentclass[aip, amsmath,amssymb, reprint, superscriptaddress]{revtex4-1}

\usepackage{graphicx}
\usepackage{dcolumn}
\usepackage{bm}

\renewcommand{\deg}{\ensuremath{{}^\circ}}

\newcommand{\etal}{\emph{et al.}}

\newcommand{\plane}[1]{\ensuremath{(#1)}}
\newcommand{\sub}[1]{\ensuremath{{}_\textrm{#1}}}
\renewcommand{\sup}[1]{\ensuremath{{}^\textrm{#1}}}

\newcommand{\units}[1]{~\ensuremath{\mathrm{#1}}}

\begin{document}

\preprint{AIP/123-QED}

\title{Orientation-dependent binding energy of graphene on palladium}

\author{Branden B. Kappes}
\email[Corresponding authors: ] {bkappes@mines.edu, cciobanu@mines.edu}
\affiliation{Department of Mechanical Engineering and Materials Science Program\\
Colorado School of Mines, Golden, Colorado 80401, USA}
\author{Abbas Ebnonnasir}
\affiliation{Department of Mechanical Engineering and Materials Science Program\\
Colorado School of Mines, Golden, Colorado 80401, USA}
\author{Suneel Kodambaka}
\affiliation{Department of Materials Science and Engineering\\
University of California, Los Angeles, Los Angeles, California 90095, USA}
\author{Cristian V. Ciobanu}
\email[Corresponding authors: ] {bkappes@mines.edu, cciobanu@mines.edu}
\affiliation{Department of Mechanical Engineering and Materials Science Program\\
Colorado School of Mines, Golden, Colorado 80401, USA}

\date{\today}

\begin{abstract}
Using density functional theory calculations, we show that the
binding strength of a graphene monolayer on Pd(111) can vary between
physisorption and chemisorption depending on its orientation. By
studying the interfacial charge transfer, we have identified a
specific four-atom carbon cluster that is responsible for the local
bonding of graphene to Pd(111). The areal density of such clusters
varies with the in-plane orientation of graphene, causing the
binding energy to change accordingly. Similar investigations can
also apply to other metal substrates, and suggests that physical,
chemical, and mechanical properties of graphene may be controlled by
changing its orientation.
\end{abstract}

\maketitle

Since its isolation in 2004,\cite{Novoselov2004Science} graphene
has attracted a great deal of attention because of its unique
physical, chemical, and electronic
properties.\cite{Altenburg2010PRL, Donner2009JCP, Kwon2009NanoLett,
Lee2008Science}  A remaining technological hurdle that prevents the
introduction of graphene into specific applications is a basic,
predictable understanding of the interaction of graphene with
supporting substrates.\cite{Giovannetti2008PRL} The sensitivity of
graphene to the substrate material and surface orientation is well
established, but even relative azimuthal orientation of graphene on
these substrates can affect its properties.\cite{Starodub2011PRB,
VazquezdeParga2008PRL,Coraux2008NanoLett, Murata2012PRB} These
recent works address manifestations of the interactions between
graphene and its host substrate, showing that these interactions actually
vary with the local environment, that is, the chemical or electronic
environment in the immediate vicinity of certain carbon
atom(s).\cite{Wang2010ACSNano} Careful bookkeeping of the local
structures (and the associated interactions) present within a single
spatial period of a (moir\'{e}) superstructure can be compared to
variations in the global properties, such as binding energy, to
deduce the impact of the orientation of the graphene layer.

In this letter, we study monolayer graphene on Pd(111) for a range
of in-plane orientations, and show that the atomic-level
carbon--palladium registry affects the binding to the substrate
through changes to molecular orbitals at the graphene--metal
interface. Furthermore, we uncover a link between the interfacial
binding strength and the areal density of certain four-atom clusters
centered atop Pd atoms in the first substrate layer. This link
provides fundamental insight into the physical origin of the
orientation dependence of the binding energy, and can be exploited
for understanding the interaction of graphene with other substrates
as well.

\begin{figure}[bth]
\begin{center}
\includegraphics[width=7cm]{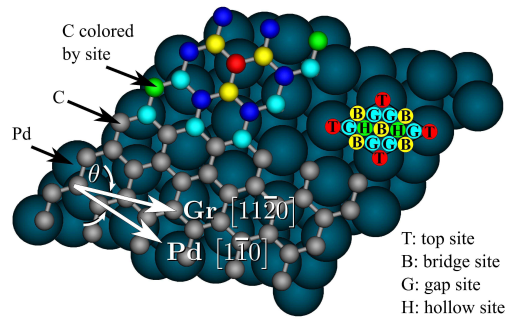}
\caption{(Color online) Possible occupancy sites for carbon atoms on
Pd(111): T (top, C atom atop a Pd atom), B (bridge, C atom directly
above the midpoint of a surface Pd-Pd bond), H (hollow, C atom above
the hollow site), and G (gap, C atom at the center of the smallest
TBB triangles).} \label{Fig1-sites}
\end{center}
\end{figure}

In order to study the orientation dependence of the binding energy,
we start by constructing perfectly periodic moir\'e patterns of
graphene on Pd(111). All moir\'e patterns are incommensurate when
using the equilibrium lattice constants of graphene and Pd, so we
have ensured commensurability at each orientation $\theta$ (defined
as in Fig.~\ref{Fig1-sites}) by applying small strains to the
graphene lattice. We have performed density functional theory (DFT)
calculations of graphene domains having orientations $\theta$ between 0
and 30\deg\ with respect to the substrate; other angular domains
reduce to $\leq\theta<30$\deg\  through rotational symmetry. For
convenience, we have chose four different orientations, $\theta$ = 5.7,
10.9, 19.1, and 30.0\deg, the first three of which have been
experimentally identified in Ref.~\onlinecite{Murata2010APL}; the
30\deg\ orientation helps compare our results with those of
Giovannetti {\em et al}.\cite{Giovannetti2008PRL} To understand the
impact of atomic registry on the graphene-Pd(111) binding, each
carbon atom is designated as a top (T), bridge (B), gap (G), or
hollow (H) site, as shown in Figure~\ref{Fig1-sites}, based on its
position relative to the substrate.

Total energy calculations were performed using the Vienna Ab-Initio
Simulation Package~\cite{Kresse1996PRB, Kresse1994JPCM} with
ultrasoft pseudopotentials in the local density
approximation,\cite{Perdew1981} with a 286\units{eV} plane-wave
energy cutoff, and a 12~\AA\ vacuum thickness. The Brillouin zone
was sampled using a Monkhorst-Pack grid--\mbox{$7 \times 7 \times
1$} (5.7\deg), \mbox{$9 \times 9 \times 1$} (10.9\deg), \mbox{$17
\times 17 \times 1$} (19.1\deg), and \mbox{$35 \times 35 \times 1$}
(30.0\deg)--with more than $100\ \textrm{points}/$\AA$^{-1}$ along
each reciprocal lattice vector. The substrate was modeled with four
Pd\plane{111} layers (lowest two kept fixed during relaxation), with
the exception of the 5.7\deg\ system, which was modeled with three
Pd layers due to computational limitations; we have tested that
binding energy trends obtained with 3 or 4 substrate layers are
indeed consistent with one another. We have computed the binding
energy per unit area $E_b$ via
    \begin{equation}
    E_b = (E\sub{GrVPd} - E\sub{GrPd})/A,
    \end{equation}
where $E\sub{GrVPd}$ is the total energy of a system in which the
graphene sheet is far from the Pd surface, and $E\sub{GrPd}$ is the
total energy of a system where the graphene layer was relaxed on the
substrate. We express the area $A$ in carbon atoms rather than
\AA\sup{2} ($1\units{C} = \sqrt{3}a^2/4$, where $a$ is the lattice
constant of graphene).

Binding energy is a fundamental property of epitaxial graphene
systems, affecting not only its structural and mechanical stability,
but the electrochemical properties as well. While calculations of binding energy
have become commonplace, our goal here is to correlate it with
simple geometric features that are present in the moir\'e
superstructures characteristic to various orientations. Seeking such
correlation involves, in order: (a) identifying the possible
locations (coincident sites) that a carbon atom in graphene can
occupy with respect to the substrate, (b) tracking the populations
of these coincident sites ({\em i.e.,} their areal densities), or of
groups of them, as functions of the orientation of the graphene
sheet, and (c) finding which of these tracked populations
depend on the orientation angle in a manner similar to the binding
energy. Following this plan, we start by selecting four distinct
types of coincident sites to describe the positions of a carbon atom
relative to the Pd surface--top (T), bridge (B), hollow (H), and gap
(G), as described in Figure~\ref{Fig1-sites}. To avoid counting
ambiguities when tracking site populations, we have chosen to
identify the regions for which coincident sites are populated by
site-centered circles that touch without overlapping, as shown in
Fig.~\ref{Fig1-sites}]. This convention ensures that 90.7\% of the
substrate surface is counted; atoms within the remaining
9.3\% of the surface (colored dark blue in Fig.~\ref{Fig1-sites}) are disregarded.

\begin{figure}[bth]
\begin{center}
\includegraphics[width=7cm]{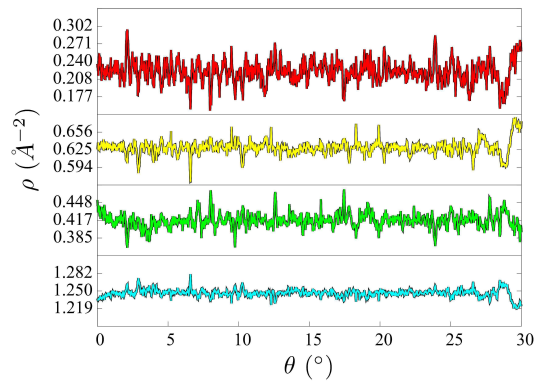}
\caption{(Color online) The areal density of the individual types of
sites occupied by the C atoms of the graphene sheet as a function of
orientation.} \label{Fig2-deviation}
\end{center}
\end{figure}

Tracking changes in coincident site population requires a fixed
reference, and we have selected this reference to be the expected
fraction of carbon atoms that would lie at a certain coincident site
under the assumption of random distribution of carbon atoms.  Placed
at random, $1/12$ of carbon atoms lie on top sites and $1/4$, $1/6$,
and $1/2$ at bridge, hole, and gap sites, respectively.
Figure~\ref{Fig2-deviation} shows the variation in the population of
each coincident site as a function of angle. Although a given
population of single-type coincident site may vary by as much as
45\% with respect to its reference, no clear trends emerge for the
variation of any coincident site population with the orientation
$\theta$ (Fig.~\ref{Fig2-deviation}). On the other hand, as we show
below, the binding energy of graphene monolayer on Pd {\em varies
monotonically with the orientation angle}: therefore, no correlation
exists with between binding energy and populations of any of the
single coincident sites, T, B, H, or G.

\begin{figure}[bth]
\begin{center}
\includegraphics[width=7cm]{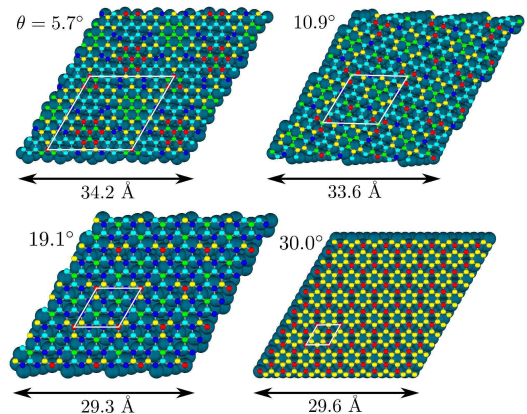}
\caption{(Color online) Graphene on Pd\plane{111} oriented at
$\theta = 5.7\deg$, 10.9\deg, 19.1\deg, and 30.0\deg,  with the
carbon atoms colored according to the site they occupy (defined in
Fig.~\ref{Fig1-sites}). The moir\'e surface unit cells are outlined
in white. Despite the nearly random distribution of top, bridge,
hole, and gap sites, we note the periodic repetition of distinctive
{\em clusters of sites} such as TB\sub{3} [three bridge site carbons
(yellow) surrounding a top site (red)], TH\sub{3} (three green,
central red), BG\sub{3} (three cyan, central yellow), HT\sub{3}
(three red, central green), HH\sub{3} (three green, central green),
HG\sub{3} (three cyan, central green), GB\sub{3} (three yellow,
central cyan), and GG\sub{3} (three cyan, central cyan).}
\label{Fig3-moires}
\end{center}
\end{figure}

\begin{figure}
\begin{center}
\includegraphics[width=7cm]{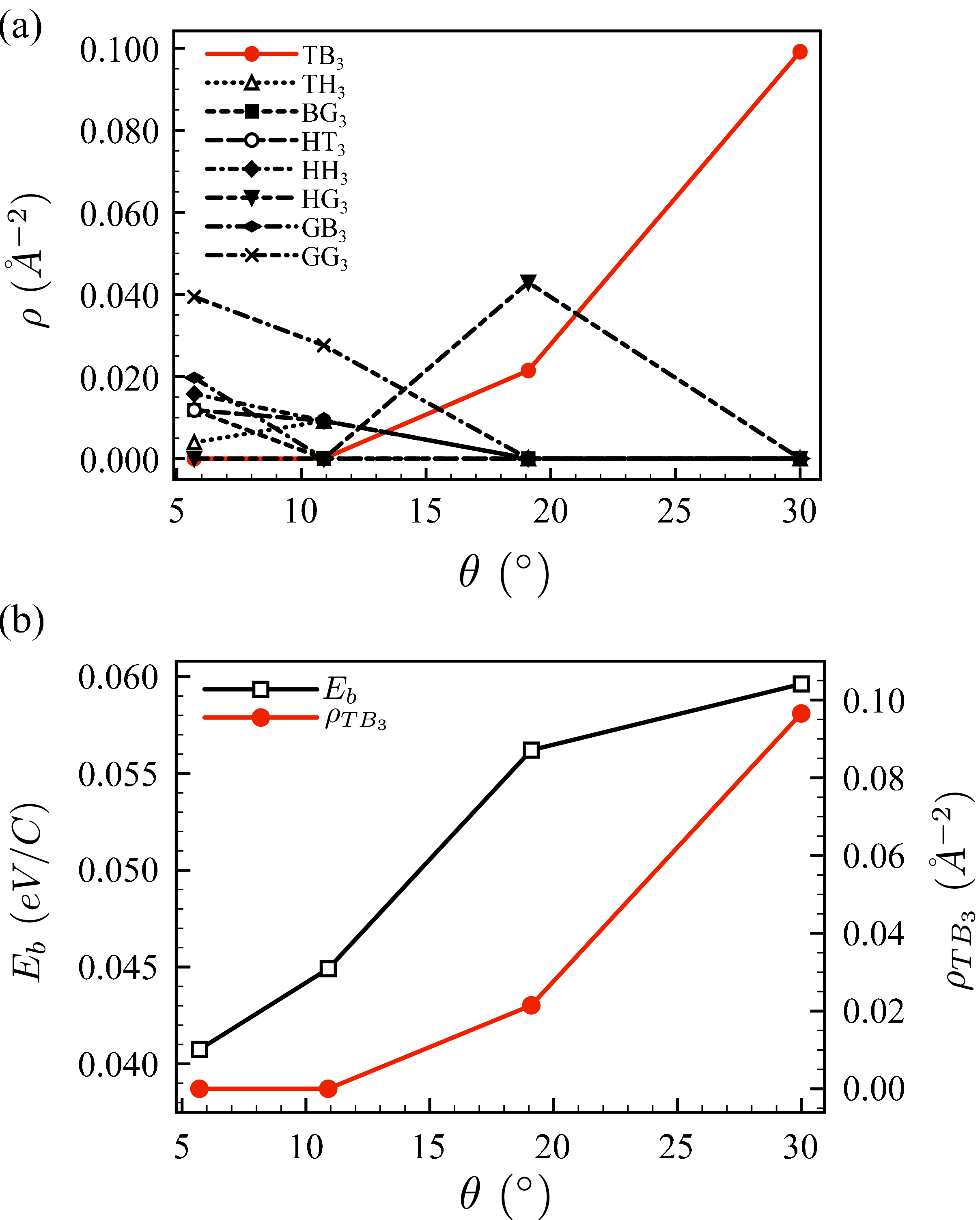}
\caption{(Color online) (a) Density of different types of 4-site
clusters for the orientations of $\theta = 5.7\deg$, 10.9\deg,
19.1\deg, and 30.0\deg. (b) The binding energy $E_b$ of graphene on
Pd(111) increases as a function of orientation $\theta$. The only
cluster whose areal density increases monotonously with $\theta$ is
TB$_3$, which we show to be responsible for the bonding of graphene
to the substrate. } \label{Fig4-binding}
\end{center}
\end{figure}

A simple inspection of the moir\'{e} patterns corresponding to the
orientations $\theta$ studied here (Figure~\ref{Fig3-moires})
suggests that coincident sites are occupied in a coordinated
manner, which is expected since the carbon atoms that occupy them are part of the same
graphene lattice (refer to examples given in the caption to Fig.~\ref{Fig3-moires}). Therefore,
we proceed to consider nearest neighbors of a given (central) site, along with
that site, in order to create an inventory of clusters that can occur on the substrate.
We have identified these clusters
by a two-letter abbreviation; for example, HG\sub{3} is a
hollow-site (H) carbon surrounded by three gap-site (G$_3$) nearest
neighbors. Eight such clusters are represented in the moir\'{e}
patterns shown in Fig.~\ref{Fig3-moires}, and we have determined the
areal density as a function of angle $\theta$ for all of them
[Fig.~\ref{Fig4-binding}(a)]. As seen in Fig.~\ref{Fig4-binding}(b),
only one of these four-atom clusters, TB$_3$, has a density that
varies with the orientation angle in the same way in which the
binding energy does. At small orientations, the graphene sheet
physisorbs to the Pd\plane{111} substrate with a binding energy of
only 41\units{meV/C}. As the sheet is rotated and the number of
TB\sub{3} clusters increases, so does the binding energy, increasing
to 73\units{meV/C} at $\theta=30$\deg\ where the density of
TB\sub{3} clusters reaches its maximum of $9.65 \times
10^{-2}\units{\AA^{-2}}$.

\begin{figure}[tbh]
\begin{center}
\includegraphics[width=7cm]{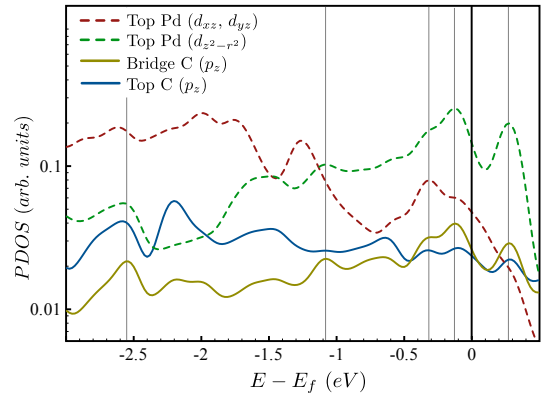}
\caption{(Color online) Site and angular momentum projected density
of states (PDOS) for the top and the bridge site carbons in a TB$_3$
cluster and for the corresponding Pd atom at the 30.0\deg\
orientation. The presence of common peaks (vertical gray lines) for
the C--$p_z$, Pd--$d_{xz}$, Pd--$d_{yz}$, and Pd--$d_{z^2-r^2}$
projected density of states is consistent with the formation of
hybridized orbitals at the interface.} \label{Fig5-pdos}
\end{center}
\end{figure}

To substantiate the link between areal density of TB$_3$ clusters
and binding energy beyond the comparison offered by
Fig.~\ref{Fig4-binding}(b), we have analyzed the site-projected
density of states (PDOS) corresponding to the atoms of the TB$_3$
cluster, as well as the electronic transfer occurring upon the
creation of the graphene--Pd interface. Given the need for very fine
Brillouin-zone sampling, the PDOS calculations were performed on the
relaxed structures with the order-\emph{N} code
SIESTA~\cite{Soler2002JPCM} using a double-$\zeta$ basis set and a
\mbox{$70 \times 70 \times 1$} Monkhorst-Pack grid.
Figure~\ref{Fig5-pdos} shows the $p_z$ states of the carbon atoms in
a TB$_3$ cluster and the $d$ states of the Pd atom beneath this
cluster. The presence of common peaks for these projected densities
of states (marked by vertical gray lines in Fig.~\ref{Fig5-pdos})
{\em below} the Fermi level and close to it indicates the formation
of hybridized, bonding orbitals between the $p_z$ states of the
carbon atoms of the cluster and the $d$-states of the Pd atom
underneath.

\begin{figure}[!tbh]
\begin{center}
\includegraphics[width=7cm]{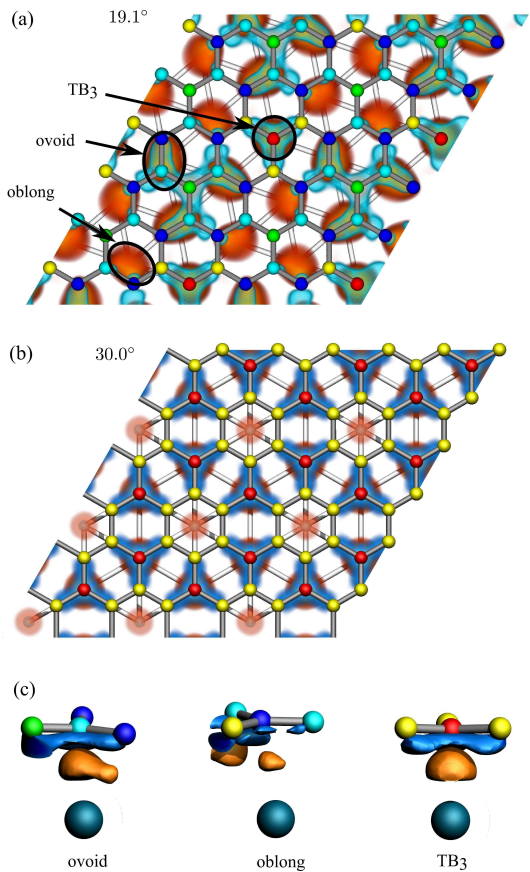}
\caption{(Color online) Electron transfer density for (a) 19.1\deg\
and (b) 30.0\deg\ orientations, showing regions of charge
accumulation (red-orange spectrum) and depletion (blue spectrum).
Three types of bonds can be identified at these orientations, marked
in (a) as ovoid, oblong, and TB$_3$. Only the TB$_3$ type exists at
$\theta=30$\deg (b), and no bonds are formed at orientations of
5.7\deg\ or 10.9\deg. Panel (c) shows side-views of the charge
transfer regions corresponding to different types of bonds
identified in (a).} \label{Fig6-chargeTx}
\end{center}
\end{figure}

Electronic transfer calculations also reveal a clear signature of
the bonding between graphene and Pd for a certain range of the
orientations. At small angles ($\theta \leq 10.9$\deg), there is no
significant charge transfer, consistent with the low binding energy
values shown in Fig.~\ref{Fig4-binding}(b). However, for
$\theta=19.1$\deg\ and 30\deg\ the electron transfer becomes
significant, as shown in Figs.~\ref{Fig6-chargeTx}(a) and (c): this
transfer amounts to the formation of chemical bonds (occupied
bonding orbitals) between graphene and substrate, which are the
physical origin of the binding energy increase computed at large
angles $\theta$ [Fig.~\ref{Fig4-binding}(b)]. In
Figure~\ref{Fig6-chargeTx}(a), we can identify three bonding
orbitals by their shapes: the axially--symmetric TB\sub{3} bond, the
ovoid bond that lies below the $\sigma$-bond between two carbon
atoms, and the oblong bond where adjacent carbon sites are not
directly atop a first-layer Pd atom. The ``strengths'' of the bonds
formed, as estimated by  the increased electronic charge, $Q$, in a
certain volume,\cite{footnote:Q} are also different
with the highest corresponding to the TB$_3$ case: $Q_\textrm{TB\sub{3}} =
0.055 e> Q_\textrm{ovoid}= 0.042 e
> Q_\textrm{oblong} = 0.007 e$. This reinforces, albeit qualitatively,
our earlier observation that the combined effect of the top and
bridge-site neighboring carbons is responsible for bonding. Unlike
the case of the 19.1\deg\ system where three types of bonds are
found, in the case of 30\deg\ only TB$_3$ is present
[Figure~\ref{Fig6-chargeTx}(b)]. The TB\sub{3} bonds are $\sim$24\%
weaker for $\theta=30$\deg\ than their 19.1\deg\ counterparts, but
are present on $2/3$ of the surface Pd atoms which accounts for the
the higher binding energy.

When bound strongly to a substrate, graphene tends to adopt one
predominant orientation, as seen on Ru(0001)~\cite{Jiang2009JCP} and
Ni(111).\cite{Starodub2011PRB} On substrates with weaker binding,
including Pd(111),\cite{Murata2010APL} Pt(111),\cite{Sutter2009PRB,
Gao2011APL} and Ir(111),\cite{Loginova2009PRB} multiple azimuthal
orientations are observed, and are readily identified by the
presence of moir\'e structures with different spatial periodicities
$L$. Recently, Merino \etal~\citep{Merino2011ACSNano} have
catalogued structures of graphene on Pt(111) into minimum-strain
``phases'', characterized by their orientation-dependent spatial
periodicities. It is worth noting that when applied to the graphene
on Pd(111) system, the Merino {\em et al.} model describes well the
structures we examined here: 5.7\deg, $L=17.12$~\AA\ (same as in
Ref.~\onlinecite{Merino2011ACSNano}, $\zeta$ phase); 10.9\deg,
$L=11.20$~\AA\  (10.75~\AA\ in Ref.~\onlinecite{Merino2011ACSNano},
$\kappa$ phase); 19.1\deg, $L=7.33$~\AA\ (7.25\AA\ in
Ref.~\onlinecite{Merino2011ACSNano}, $\beta$ phase); and 30\deg,
$L=4.89$~\AA\ (5.00\AA\  in Ref.~\onlinecite{Merino2011ACSNano},
$\alpha$ phase). The most stable phases (orientations) are those
with stronger binding, {\em i.e.,} $\alpha$ and $\beta$, as shown
before in Fig.~\ref{Fig4-binding}(b).

In summary, we have shown that the binding energy of graphene on
Pd(111) depends on its orientation, and that this dependence arises
not from the population of any single coincident site, but from
coincident site four-atom clusters (one top-site C atom surrounded
by three bridge-site carbons). The 78\% increase in binding energy,
from 41\units{meV/C} to 73\units{meV/C}, emphasizes the sensitivity
of this property to the carbon--palladium coincidence. The approach
presented here can be applied to other metal substrates as well, in
order to see if equally simple clusters (and which ones) are
responsible for controlling the binding strength between
physisorption and chemisorption.

{\em Acknowledgments.} BBK, AE, and CVC gratefully acknowledge the
support of National Science Foundation through Grant Nos.
CMMI-0825592, CMMI-0846858, and OCI-1048586. SK gratefully
acknowledges funding from the Office of Naval Research, ONR award
no. N00014-12-1-0518 (Dr. Chagaan Baatar). Computational resources
for this work were provided by the Golden Energy Computing
Organization at Colorado School of Mines.


\begin{thebibliography}{22}
\bibitem [{foo()}]{footnote:Q}%
  \BibitemOpen
  \href@noop {} {}\bibinfo {note} {In the plot of spatial charge distribution,
  the electronic charge ``contained'' within a bond, $Q$, depends on the choice
  of the isosurface contour limits, but the relative strengths of the bonds in
  a structure can be compared when the charge contained in each is measured
  using the same contour limits. Studies of the spatial inflections of the
  density of charge transfer lead us to an isosurface limit level of $1 \times
  10^{-5}e$.}\BibitemShut {Stop}%
\end{thebibliography}
%

\end{document}